\documentstyle[referee]{mn}
\onecolumn
\input{epsf}
\title{Line-of-Sight Anisotropies within X-ray Luminous Cluster Samples}
\author[Miller, Ledlow, \& Batuski]{Christopher J. Miller $^1$, Michael J. Ledlow$^2$, David J. Batuski$^1$\\
$^1$Department of Physics and Astronomy, University of Maine, 
Orono, Maine 04469-5709, U.S.A.\\
$^2$Institute for Astrophysics, University of New Mexico, Albuquerque, New Mexico 87131, U.S.A.\\}

\pagerange{000,000}

\begin{document}

\maketitle

\begin{abstract}
We present the results from two-point spatial correlation
analyses on X-ray confirmed northern Abell clusters.
The cluster samples are subsets of a volume-limited ROSAT All-Sky
Survey study of 294 $R \ge 0$ Abell clusters of which 240 are X-ray 
luminous. This large number of clusters has allowed for magnitude- and
volume-complete samples to be analysed according to richness and
X-ray luminosity. For $R \ge 1$ clusters, we find $r_0 = 22h^{-1}$Mpc
and $\gamma = -1.7$, which is consistent with previous
analyses of visually selected $R \ge 1 $ Abell clusters. We also find no
indications of line-of-sight anisotropies within the $R \ge 1$ clusters.
For $R \ge 0$ clusters, we find $r_0= 17.5h^{-1}$Mpc (and $\gamma = -1.8$) which
is considerably lower than recent determinations of the correlation length for
similar $R \ge 0$ X-ray
bright cluster samples (e.g. the X-ray Brightest Abell Cluster sample (XBACs)
with $21 \le r_0 \le 26 h^{-1}$Mpc
and the RASS1 X-ray cluster sample with $r_0 \sim 23 h^{-1}$Mpc).
All of the $R \ge 0$ X-ray confirmed samples, including the XBACs and RASS1 clusters
show line-of-sight anisotropies.
Since X-ray emissions confirm a cluster's reality, we
conclude that these line-of-sight anisotropies
are not the result of spuriously selected clusters.
These results conflict with past conclusions that the correlation length
of $R \ge 0$ Abell clusters is artificially enhanced due to anisotropies
caused by spurious cluster selection.
We also examine
a magnitude- and volume-complete sample of $R \ge 1$ Abell clusters for
the depedence of $r_0$ and $\gamma$ on X-ray luminosity, and find
no evidence for $r_0$ to grow with increasing X-ray luminosity thresholds.
This is contrary to similar L$_x$ vs. $r_0$ analyses of the RASS1 and XBACs cluster samples.
We describe selection effects within the flux-limited XBACs and RASS1 samples and suggest how
they can affect both the size of the correlation length and its dependence
on $L_x$.

\end{abstract}

\section{Introduction}

The two-point spatial correlation function is 
used to describe the scale of clustering within
discrete datasets. Both galaxies and clusters of
galaxies have a functional power-law 
form for the correlation
function,
$\xi(r) = (r/r_0)^{\gamma}$.
The slope and amplitude of this power law is rather well-defined
for galaxies to be $r_0 = 5h^{-1}$Mpc and $\gamma = -1.8$
({\it e.g} Willmer {\rm et al.} 1998 and references therein). 
For galaxy clusters, the slope has been established
at $-2.0 \le \gamma \le  -1.8$, but 
the value for $r_0$ has
been a matter of much debate.
The majority of cluster-cluster spatial correlation analyses have
been based on the visually \lq\lq scanned\rq\rq~
Abell and ACO catalogs (Abell 1958; Abell, Corwin and Olowin 1989)
and the machine scanned Automatic Plate Measuring (APM) Facility
Cluster Survey (Maddox {\rm et al.} 1990a,b). The correlation length
for the visually selected clusters is $\sim 20-25 h^{-1}$Mpc with
positive correlations out to separations of $\sim 50 h^{-1}$Mpc
({\it e.g.} Miller {\rm et al.} 1999 and references therein). 
However, the clusters selected through machine scanning have 
$r_0 \sim 15h^{-1}$Mpc and little positive correlation beyond
25$^{-1}$Mpc (Efstathiou {\rm et al.} 1992; Dalton {\rm et al.}
1994). 

The large differences between the above determinations of $r_0$ for clusters have been
explained in either of two ways:
\begin{enumerate}
\item{The optically selected clusters suffer from spurious cluster selection.
This observational selection bias occurs when two clusters are near
each other on the plane of the sky, but separated by a large distance radially.
When this occurs, the richness of either the foreground or background cluster may be
artificially enhanced due to projection effects. This line-of-sight selection bias
creates false spatial correlations at larger separations, which in turn inflates
$r_0$ ( e.g. Sutherland 1988; Efstathiou {\rm et al.} 1992).
We point out that a substantial number of clusters missed in a non-random
sytematic matter during the visual selection
process can also give rise to this effect.}
\item{The value of $r_0$ is dependent on the mean cluster number density ($n_c$) of the sample,
\begin{equation}
r_0 = 0.4n_c^{-1/3}.
\end{equation}
In this case, the APM clusters should have a smaller
correlation length since their number density is nearly four times that of
$R \ge 1$ Abell
clusters (Bahcall \& West 1992; Bahcall \& Cen 1994).}
\end{enumerate}

While both of the above solutions seem plausible and
explain (and/or correct) the value of $r_0$, both solutions
have also been shown to be flawed.  Line-of-sight anisotropies within
the Abell and ACO catalogs have been examined in detail by Miller {\rm et al.}
(1999) who find that only $\sim 10\%$ of clusters in the ENACS (Katgert {\rm et al.}
1996) and MX (Slinglend {\rm et al.} 1998) surveys show strong background/foreground
contaminations. In addition, Miller {\rm et al.} find $r_0 \sim 22h^{-1}$Mpc  for
$R \ge 1$ clusters both before and after
removing these contaminated clusters from the analysis. They also show that
the minimal 
anisotropy present in the $R \ge 1$ subset of clusters is similar in scale to that of
the APM clusters. 
Miller {\rm et al.} conclude that projection effects and line-of-sight anisotropy
are not large problems for $R \ge 1$ Abell/ACO clusters and do not artificially
enhance $r_0$.

On the other hand, the density dependence on the correlation length was determined
empirically and ultimately depends on the accurate evaluation of $r_0$ and the
mean cluster density for mulitple
samples. While many of the currently available cluster
datasets
have mean densities $\sim 1 \times 10^{-5}h^{3}$Mpc$^{-3}$ or greater, until
recently, only the richest ($R \ge 1$)
Abell clusters have provided $r_0$ for datasets with
densities  $\sim  \times 10^{-6}h^{3}$Mpc$^{-3}$. 
Croft {\rm et al.} (1997) constructed a catalog of very rich APM
clusters with a mean number density of $\sim 1 \times 10^{-6}h^{3}$Mpc$^{-3}$
and find $r_0 = 21h^{-1}$Mpc which is contrary to the expected result from 
Equation (1). 
Unfortunately, we do not have a statistically significant
determination of $r_0$ for $R \ge 2$ Abell clusters (with 
$\bar{n} \sim 1 \times 10^{-6}h^{3}$Mpc$^{-3}$), although results
from Peacock \& West (1992) suggest that the correlation length may be as high as
$r_0 = 45h^{-1}$Mpc.
With only two very rich samples studied so far, the
$r_0 \propto  n_c^{-1/3}$ relation lacks strong observational support
for densities less than $10^{-5}h^{3}$Mpc$^{-3}$.
In addition to the observational analyses,
both Croft \& Efstathiou (1994) and Eke {\rm et al.} (1996) find that the
density dependence on the correlation length is at best very weak in N-body
simulations.

In this work, we will examine magnitude-limited and volume-limited
samples of Abell clusters which are also X-ray luminous. The problems of projection
effects and spurious cluster selection are minimized in X-ray bright clusters,
allowing a more reliable determination of the amplitude and slope of the
two-point spatial correlation function. We would expect that any
line-of-sight anisotropies present in optically limited samples
would not be present in X-ray confirmed cluster samples.  However, while 
the X-ray emission of the intracluster gas will confirm the reality of
a catalogued cluster, this approach provides no information on clusters missed by Abell (1958)
in his visual search. With the recent work on X-ray selected cluster catalogs
( e.g. Ebeling {\rm et al.} 1998; Vikhlinin {\rm et al.} 1998), we can
expect future two-point spatial correlation analyses that would include
any optically missed galaxy clusters as well (see De Grandi {\rm et al.} 1999).

\section{Data and Methods}
The X-ray luminosities and their associated
uncertainties were taken from Voges, Ledlow, Owen
and Burns (1999). Voges {\rm et al.} studied RASS
(ROSAT All-Sky Survey) data for a volume limited ($z \le 0.09$)
sample of 294 $R \ge 0$ Abell clusters.
These clusters have the
following criteria: log$_{10}$ N$_H < 20.73$ (roughly corresponding to $|b| \ge 25^{\circ}$ which
we apply as a cut-off),
$z \le 0.09$ and $\delta \ge -27^{\circ}$. 
All of the X-ray luminous Abell clusters
used in this work
have measured redshifts.
Voges {\rm et al.} found that 84\% of the $R \ge 0$ Abell clusters in their
volume-limited sample were X-ray luminous. The majority of the clusters that showed
no X-ray emissions 
were $R = 0$ clusters.

We apply
magnitude and richness constraints to the Voges {\rm et al.} sample so that we may examine statistically
complete samples.
Specifically, we will divide the clusters into two subsets with different
richness class ranges, one with
$R \ge 0$ and the other with $R \ge 1$. For each of these richness-limited subsets, we will
look at volume-limited ($z \le 0.09$) cluster samples with and without appropriate magnitude
limits so that they may be considered statistically complete for comparision to other
such analyses.
For the $R \ge 0$ clusters, we will use a magnitude limit of m$_{10} \le 16.5$ and for
$R \ge 1$ clusters we will use  m$_{10} \le 16.8$. Postman, Huchra, \& Geller (1992-hereafter PHG)
presented correlation analyses for the
magnitude-complete sample of $R \ge 0$, m$_{10} \le 16.5$ clusters. Miller {\rm et al.} (1999)
presented correlation analyses for the magnitude-complete sample of  $R \ge 1$, m$_{10} \le 16.8$ clusters.
Both of these samples were based on visually selected clusters. 

The dependence of the X-ray luminosity on cluster mass
($L_x \propto M^{p}$) has been well established 
both analytically and numerically (Bertschinger 1985;
Evrard \& Henry 1991;  Navarro,
Frenk, and White 1995, Ledlow {\rm et al.} 1999). Figure 1 shows the observational
results of the $L_x - M$ relation
using 42 cluster virial-masses from
Girardi {\rm et al.} (1998).
An outlier-resistant linear-fit to the data in Figure 1 produces
$L_x \propto M^{2.38\pm{1.27}}$.
The errors bars on each data point are 
$1 \sigma$ Poisson in the virial mass determination 
as provided by Girardi {\rm et al.} and
$1 \sigma$ in the X-ray luminosity determination
from Voges {\rm et al.} 
This observational result for $p$ is consistent with simulations by
Ledlow {\rm et al.}, but is significantly steeper than the self-similar
scaling laws of Kaiser (1986) which predicts L$_x \propto$ M$^{4/3}$.
The analytical, numerical, and observational evidence
for $L_x \propto M^{p}$ suggests that we should also examine the
cluster-cluster correlation function for dependence
on X-ray luminosity. Such a dependence has been found in
the XBACs and the RASS1 clusters, although both results
can not be considered statistically significant (Abadi {\rm et al.} 1998;
Borgani {\rm et al.} 1999;
Moscardini {\rm et al.} 1999).

We use the following estimator derived in Hamilton (1993) for
the determination of the correlation function: 
\begin{equation}
\xi(r) = \frac{DD(r) \times RR(r)}{DR(r)^2} -1,
\end{equation}
where $DD$, $RR$, and $DR$ are the data-data, random-random and 
data-random paircounts respectively
with separations between $ r - \frac{\Delta r}{2}$ and $ r + \frac{\Delta r}{2}$.
We refer the reader to Hamilton (1993) and Landy \& Szalay (1993)
for an analytical analysis of the 
estimator.
Compared to previous estimators (Bahcall \& Soneira 1983;
PHG), this one is proposed
to be less affected by
uncertainties in the mean number density where separations are
large and $\xi$ is small.
Recently, Ratcliffe {\rm et al.} (1998) used N-body simulations to show that
Equation (2) provided the most accurate results when compared
to other estimators. 

The random paircounts (DR, RR) are evaluated by averaging over 400
catalogs generated with the same number of pseudo-clusters as the
sample under consideration. The angular coordinates in these catalogs
are randomly assigned
with the same boundary conditions as the survey.
While cluster X-ray emission is not entirely hidden due to galactic
obscuration, the clusters themselves were catalogued
optically (note: corrections to the X-ray luminosities were made by Voges {\rm et al.} (1999)
to account for galactic absorption),
therefore, the known selection bias in $b$ is carried
into all subsets of the original catalog. To account for this, we apply a
latitude selection function;
\begin{equation}
P(b) = 10^{\alpha(1-csc|b|)},
\end{equation}
with $\alpha = 0.32$.
The redshifts assigned to the random catalog points are 
selected from the observed data after being smoothed
with a Gaussian of width 3000 km s$^{-1}$. This technique
corrects for radial density gradients on small scales in the
observed distribution.
Distances to all clusters were calculated assuming a Friedman universe
with $q_0$ = 0 and $H_0$ = 100 km s$^{-1}$ Mpc$^{-1}$. 

\section{Results}
 
The results for power law fits to the cluster-cluster
two-point spatial correlation function are given in
Table 1 and Figures 2 and 3.
It is important to note that samples 2 and
4 are volume-limited only and are 
incomplete in magnitude. Volume-limited surveys are not well suited
for comparison to other works, since some very dim but nearby clusters will be
added to the volume over time. For instance, since the time the Voges {et al.}
volume-limited optically complete sample was defined, an additional 86 clusters
have since been observed that are within z = 0.09. This is nearly a 30\% increase
in only a few years time. On the other hand, a magnitude-limited sample will
always contain the same number of clusters (if complete). 
The error bars in Figure 2 are
determined from
\begin{equation}
\delta\xi = \frac{(1+\xi)}{\sqrt{DD}}.
\end{equation}
However, we note that Croft \& Efstathiou (1994) have shown that this underestimates
the true error by a factor of
$1.3 \rightarrow 1.7$.

There are significant differences in $r_0$ and $\gamma$ between the 
$R \ge 0$ clusters and the $R \ge 1$ clusters.
The two most important aspects of these results are:
\begin{enumerate}
\item{$\gamma$ and $r_0$ for sample 1 differ significantly from those
of PHG who find $r_0 = 20.0h^{-1}$Mpc and
$\gamma = -2.5$  for
an optically selected (m$_{10} \le 16.5$, $z \le 0.08$) complete sample of
$R \ge 0$ clusters. If we constrain the slope for sample 1 to that of
PHG, we find $r_0 = 13.5h^{-1}$Mpc
which differs from their results by 2$\sigma$.}
\item{The results for samples 3 and 4 confirm a large
correlation length for $R \ge 1$ clusters as seen previously by Bahcall \& Soneira (1983),
PHG, Peacock \& West (1992) and Miller {\rm et al.} (1999),
using visually selected Abell clusters.}
\end{enumerate}

\begin{table}
\begin{center}
\caption{Results for the Power-Law Fits of $\xi$} 
\begin{tabular}{cccccc}
\hline
Sample &
Size  &
$R$ &
m$_{10}$ &
$\gamma$  &
$r_0$ ($h^{-1}$Mpc)  \\
\hline
1 & 189 & $\ge 0$ & $\le 16.5$ & $-1.78\pm{0.20}$ & $17.7^{+3.8}_{-4.5}$ \\
2 & 240 & $\ge 0$ & all     & $-1.87\pm{0.18}$ & $17.3^{+3.0}_{-2.7}$ \\
3 & 117 & $\ge 1$ & $ \le 16.8$ & $-1.70\pm{0.25}$ & $21.5^{+6.2}_{-7.0}$ \\
4 & 130 & $\ge 1$ & all        & $-1.68\pm{0.22}$ & $22.4^{+5.9}_{-6.8}$ \\
\hline
\end{tabular}
\end{center}
\end{table}

\begin{table}
\begin{center}
\caption{Results from Other X-ray $R \ge 0$ Cluster Surveys}
\begin{tabular}{ccccc}
\hline
Reference &
Cluster Sample &
N$_{clusters}$ &
{$r_0 (h^{-1}Mpc)$} &
{$\gamma$} \\ 
\hline
Nichol  {\it et al.} 1994 &Abell & 67 & 16.1  $\pm{3.4}$ & -1.9  $\pm{3.4}$ \\
Abadi {\it et al.} 1998  & XBACs & 248 & $21.1^{+1.6}_{-2.3}$ & -1.92 \\
Moscardini {\rm et al.} 1999 & RASS1 & 130 & $22.7 \pm{3.6}$ & $-2.08^{+0.43}_{-0.51} $ \\
Borgani {\rm et al.} 1999 & XBACs & 203 & $26.0 \pm{4.5}$ & $-2.00 \pm{0.4}$ \\
This work &  Abell & 240 & $17.3^{+3.0}_{-2.7}$ & $-1.87\pm{0.18}$ \\ 
\hline
\end{tabular}
\end{center}
\end{table}

\begin{table}
\begin{center}
\caption{Correlation function as a function of increased luminosty cut-off.}
\begin{tabular}{cccccc}
\hline
Sample &
Size  &
$L_x \times10^{43}$ &
$\gamma$  &
$r_0$ & 
$r_0$ for $\gamma = -1.8$ \\
 & & $h^{-2}$ergs s$^{-1}$ & &
($h^{-1}$Mpc) &
($h^{-1}$Mpc) \\
\hline
5 & 103 & $\ge 0.14$ & $-1.88\pm{0.32}$ & $22.2^{+10.5}_{-11.1}$ & 23.1 \\
6 &  93 & $\ge 0.28$ & $-1.99\pm{0.45}$ & $17.4^{+11.2}_{-12.4}$ & 19.2 \\
7 &  80 & $\ge 0.42$ & $-2.30\pm{0.49}$ & $17.2^{+11.6}_{-13.0}$ & 20.8 \\
8 &  71 & $\ge 0.56$ & $-2.52\pm{0.60}$ & $15.0^{+13.0}_{-13.4}$ & 19.0 \\
\hline
\end{tabular}
\end{center}
\end{table}
We can compare the results for $r_0$ to those predicted by the
average number densities, $\bar{n}$, for samples 1 and 3.
PHG report $\bar{n} = 1.2\times10^{-5}h^{3}$Mpc$^{-3}$
for $R \ge 0$ Abell clusters ({\it i.e.} sample 1), while
Miller {\rm et al.} (1999) report $\bar{n} = 6.6\times10^{-6}h^{3}$Mpc$^{-3}$
for $R \ge 1$ Abell clusters. Using these densities, Equation (1) predicts
$r_0 = 17.5h^{-1}$Mpc and $r_0 = 21.3h^{-1}$Mpc for samples 1 and 3
respectively. If there are a substantial number of spuriously selected $R=0$ clusters
in the PHG sample, their calculated number density would be over-estimated. Using the methods
of Miller {\rm et al.} (1999), we calculate 
$\bar{n} = 8.68\times10^{-6}h^{3}$Mpc$^{-3}$ using only X-ray luminous Abell clusters (i.e. Sample 2),
which corresponds to $r_0 = 19.5 h^{-1}$Mpc.
These predictions are well within the $1\sigma$ uncertainties on $r_0$
given in Table 1. 
In Table 2, we list other two-point correlation function results for X-ray bright
cluster samples that include $R \ge 0$ clusters.
Our results are most similar to those of Nichol {\rm et al.} (1994).
We discuss possible explanations for the differences in the values of $r_0$ in 
Section 3.2.

\subsection{Anistropies within X-ray Cluster Samples}
At this point, we should also examine these samples for line-of-sight
anisotropies such as those found in the PHG 
$R \ge 0$ sample (Efstathiou {\rm et al.} 1992).
Line-of-sight anisotropies have been suggested by many to be responsible for the
high value of $r_0$ found by PHG in the $R \ge 0$ clusters.
The only other work on 
anisotropies in X-ray confirmed cluster samples
was performed by Nichol {\rm et al.} (1994). Their sample of 67 $R \ge 0$
Abell clusters has little line-of-sight anisotropy and a correlation length of
$r_0 = 16h^{-1}$Mpc which is significantly lower than that found by PHG (note:
both the Nichol and PHG samples are $R \ge 0$).
Sutherland (1988) was the first to show that by
dividing the pair separation vector ($\vec{r}$)
into its line-of-sight ($\pi$) and
perpendicular-to-the-line-of-sight ($\sigma$) components, one can look
for strong correlations in $\xi(\sigma, \pi)$ where 
$\sigma$ is
small ($0-20h^{-1}$Mpc) and $\pi$ is large ($30-100h^{-1}$Mpc). 
Sutherland suggested that if such line-of-sight anisotropies exist, there must be
a substantial amount of spurious cluster selection due to
foreground/background contamination. Clearly, if the clusters are confirmed 
by their X-ray brightness (which is not affected by projection effects), then
we would expect no spurious clusters and little line-of-sight anisotropy (as
seen in the Nichol {\rm et al.} results). 

In Figures 4 and 5 we present contour plots of $\xi(\sigma, \pi)$ for
samples 1 and 3.  The bold line is
$\xi(\sigma, \pi) = 1$, indicative of relatively strong 
correlations. In an ideal sample with no line-of-sight anisotropies,
small cluster peculiar velocities and non-elongated superclusters, one might
hope to find $\xi > 1$ contours with similar extent (say 25$h^{-1}$Mpc)
in $\sigma$ and $\pi$.  Notice that the $R \ge 0$ clusters show
strong correlations for $\sigma < 20h^{-1}$Mpc and $\pi$ out to 80$h^{-1}$Mpc.
This indicates that an unexpectedly large number of clusters pairs have
small separations on the plane of the sky, while being separated by
large distances radially.
Our $R \ge 0$ subset shows the
same strong anisotropies seen in the optical sample of $R \ge 0$
clusters presented by Efstathiou {\rm et al.} (1992), even
though all clusters have been confirmed by their X-ray emission.
Therefore, while some type of anisotropy does
exist in the $R \ge 0$ subset, it is not the result of spuriously
selected clusters. 

Both of the samples that exclude $R =0$ clusters show no
indications of line-of-sight anisotropies in Figure 5.
This supports claims made by Miller {\rm et al.} (1999) that
the $R \ge 1$ subset of Abell clusters do not suffer from
serious projection contamination and line-of-sight anisotropies.
In addition, the value of $r_0$ for X-ray selected $R \ge 1$
Abell clusters is similar to that found by Miller {\rm et al.} 
using their newly enlarged sample of Abell/ACO clusters with measured redshifts.
It is important to recognize that the subsets showing the
most line-of-sight anisotropies have the lowest value for
$r_0$. This is precisely the opposite of what has been suggested
by Sutherland (1992), Efstathiou {\rm et al.} (1992), 
Dalton {\rm et al.} (1994) and Nichol {\rm et al.} (1994) among others
(see section 1).
From these results we conclude that the correlation length is affected
by the richness of the cluster sample and that Equation 1 works well
for the samples analysed here,
although we point out that a higher
richness class sample (i.e. $R \ge 2$) is needed to verify such a relation for
all richness classes. We conclude that while line-of-sight anisotropies
are present in the $R \ge 0$ samples, there is no indication that they artificially inflate
the correlation length. 

\subsection{Comparison to Other X-ray Samples}
With the recent increases in the amount of available X-ray data for
clusters, other X-ray cluster samples have also been
examined for structure using the two-point correlation function.
Specifically, the Ebeling {\rm et al.} (1996) XBACs sample
is a flux-limited survey of Abell/ACO clusters in the northern
and southern galactic hemispheres, and the De Grandi {\rm et al.} (1999)
RASS1 sample is a
flux-limited survey of clusters in the Southern hemisphere.
The two-point correlation function for the XBACs was presented by Abadi {\rm et al.}
as well as Borgani {\rm et al.}, while Moscardini {\rm et al.} have presented correlation
results for the RASS1 cluster samples. (The results for these studies are presented in
Table 2).

The cluster sample in this work differs significantly from that of the 
XBACs sample which has a disproportionate distribution of richness classes.
For instance, the XBACs sample contains 25\% $R=0$ clusters with an
average redshift of $z \sim 0.073 $, 39\% $R = 1$ with an average redshift of
$z = 0.085 $,  while the remaining $R \ge 2$ clusters have an
average redshift of $z = 0.109 $. For the Voges et al. $R \ge 0$ cluster Sample 1,
46\% are $R = 0$ clusters  with an average redshift
of $z = 0.064$, 44\% are $R \ge 1$ clusters with an average redshift of
$z = 0.067$ and the remaining $R \ge 2$ clusters have an average redshift
of $z = 0.066$. The XBACs sample is not
homogenuous in richness and includes nearby poorer clusters
and generally more distant rich clusters. In Figure 6 we present the anisotropy plot for the
XBACs sample. There is strong evidence for extreme line-of-sight anisotropy as the result
of cluster pairs with small separations on the plane of the sky and large separations
in redshift.
It is of interest to note that a large fraction ($\sim 50\%$)  of the cluster
pairs causing this anisotropy are located within a very small range
of R.A. ($0^h \le \alpha \le 3^h$) and 25\% are located within an area
of only $\sim 0.04$ steradians (corresponding to roughly 1\% of the total area
covered by the XBACs). While we cannot explain the apparent pair-selection bias in the
XBACs sample,
the disproportionate fraction of higher richness clusters in the XBACs
is a direct result
of flux-limited surveys (as shown by Bahcall \& Cen 1994 ).
Using the richness-dependence of the correlation length (Equation 1), a
large fraction of $R = 1$ and $R \ge 2$  clusters would inflate 
the correlation length as compared to a sample containing a homogeneous distribution
of richness class clusters ($R \ge 0$).
[Note: a homogeneous richness distribution of Abell and ACO clusters is complicated
by classification differences between the two catalogs (see e.g. David, Forman \& Jones 1999).]

The RASS1 sample contains
130 clusters, the majority of which (101) are Abell/ACO clusters. We examine the richness
distribution of this sample and find that 11\% are $R =0$ clusters with an average
redshift of $z = 0.066$, 22\% are $R = 1$ clusters with an average redshift of $z = 0.094$
and 32\% $R \ge 2$ clusters with an average redshift of $z = 0.104$.  
The remaining 29 clusters in the
RASS1 sample are clusters ``missed'' by Abell/ACO. Most of
these are poorer APM or Zwicky clusters while some are newly identified. The average
redshift of these clusters is $z = 0.091$. In any event,
the vast majority of clusters in this sample are Abell/ACO and many of the others
would not have met Abell's richness criteria.
In Figure 7 we present the anisotropy plot for the RASS1 cluster sample. While the
line-of-sight anisotropy is not as problematic as in the XBACs, there is still
more than in the $R \ge 1$ Abell cluster sample examined in this work. As in the XBACs case, the large 
fraction of $R \ge 1$ clusters will increase the correlation length of the RASS1 sample
(compared to the more homogeneously distributed Voges {\rm et al.} sample).

\subsection{L$_x$ and Richness Dependence on $r_0$ and $\gamma$}

From the $L_x -M$ relation shown in section 2, we are also
interested in any trend in $r_0$ and $\gamma$ with respect to
$L_x$.
We created
four magnitude- and volume-limited samples with increasing 
$L_x$ cutoffs. These samples are subsets of Sample 3 and the
results for $r_0$ and $\gamma$ are presented in Table 3.
Notice that we see no increase in $r_0$  with respect to
increased $L_x$. While at first glance it may appear as if
$r_0$ is actually decreasing as we raise the $L_x$ cutoff, this is simply the
result of a steepening slope (the final column in Table 2 lists
the value of $r_0$ when the slope is constrained to $\gamma = -1.8$).
We also examined the other three samples and found no increase in $r_0$ with
increasing $L_x$ cutoff.
These results contradict those using the XBACs and RASS1 clusters in which there
is seen a weak dependence in $r_0$ with increasing $L_x$
(Abadi, Lambas, \& Muriel 1998; Borgani {\rm et al.} 1999;
Moscardini {\rm et al.} 1999).
However, none of the results based on the XBACs and RASS1 clusters
can be considered statistically significant due to small sample sizes. Voges {\rm et al.}
have found a significant correlation between X-ray luminosity
and cluster richness (where the probability of no correlation is $< 1\times{10^{-4}}$).
Therefore,
if a cluster sample has a disproportionate richness distribution (as in both the XBACs and
the RASS1 clusters), $r_0$ will increase (due to the richness dependence on $r_0$)
as the higher $L_x$ cutoff excludes poorer
clusters.
Thus, the $r_0$ -$L_x$ dependence seen in the XBACs and RASS1 clusters
is most likely an artefact of the biased cluster
samples used in their analyses.

With the recent discovery by Loken, Melott, and Miller (1999) that X-ray cooling
flow clusters with the highest mass deposition rates are located in dense cluster
environments, we decided to examine the nearest-neighbor distribution ($nnd$)
of the Voges {\rm et al.} (1999)
X-ray luminous Abell sample for any similar correlations. Specifically, we divided
the sample into the same X-ray luminosity classes as described in Table 3 plus a class
of clusters with no detected emissions. We then
determined the average nearest-neighbor distance for each class. The clusters with
no detected X-ray emissions had the smallest distance, $<nnd> = 17.3h^{-1}$Mpc, with an
increasing $<nnd>$ for 
clusters with the highest X-ray luminosities 
($L_x > 0.56\times{10^{43}}h^{-2}$ergs s$^{-1}$ ) with a distance of $<nnd> = 19.6h^{-1}$Mpc. 
However, a K-S analysis of the $nnd$ distributions show no significant differences
among the luminosity classes examined. Thus, while a slight trend for the average
nearest-neighbor distance to increase with increasing luminosity is detected, it
is not of statistical significance.

\section{Conclusions}
We have analysed magnitude- and volume-limited samples of Abell
clusters for the amplitude and slope of the two-point spatial correlation function
and also for line-of-sight anisotropies.
We find $r_0 = 17.5h^{-1}$Mpc and $\gamma = -1.8$ for $R \ge 0$
clusters, which is consistent with the results of Nichol {\rm et al.} (1994).
However, we find that the $R \ge 0$ subset contains considerable
line-of-sight anisotropies even after all clusters have been confirmed
by their X-ray brightness. For $R \ge 1$ clusters we find $r_0 = 22h^{-1}$Mpc
and $\gamma = -1.7$ and no indications of line-of-sight anisotropy. We
conclude that (1) some type of anisotropy is present in $R = 0$
clusters, although it is not the result of spuriously selected clusters. We
suggest that this anisotropy could be caused by Abell systematically searching
for (or noticing) $R =0$ clusters only in the vicinity of richer clusters and therefore
missing a substantial number of more isolated $R=0$ clusters; 
(2) the
correlation length is not artfically inflated by line-of-sight anisotropes; 
(3) there is a cluster richness dependence on $r_0$ (or mean density) in 
cluster subsets, although it is hard to say what effect the
anisotropy in the $R=0$ clusters has; (4) there is no correlation between $r_0$ and $L_x$ for
X-ray confirmed Abell clusters.

These results confirm the value for $r_0$ by previous studies using optically
limited samples of $R \ge 1$ Abell clusters (Bahcall \& Soneira 1983;
Miller {\rm et al.} 1999). Yet at the same time, we find no evidence for
the correlation length to be artifically inflated as the result of spuriously
selected cluster.
Many researchers have advocated `corrective' techniques
for dealing with $R \ge 0$ samples of Abell clusters. These techniques typically
involve the exclusion of questionable clusters (or cluster pairs) from
correlation analyses which, in turn, lowers $r_0$ considerably
(Sutherland 1988; Efstathiou {\rm et al.} 1992). Our findings
indicate that, even after using X-ray confirmed $R =0$ Abell clusters, as well as samples
including clusters missed by Abell (the RASS1 sample),
line-of-sight anisotropies are still present.
The only X-ray cluster sample that shows no such anisotropy is
the $R \ge 1$ Abell cluster subset.
We find little difference in $r_0$ between visually selected clusters ($r_0 \sim 22h^{-1}$Mpc, Miller
{\rm et al.} 1999) and X-ray confirmed $R \ge 1$ Abell clusters ($r_0 = 22h^{-1}$Mpc, 
presented in this work) for similar slopes ($-1.8 \le \gamma \le -1.7$). 

The correlation length for rich clusters of galaxies
has been debated for well over a decade. 
During that time, no other cluster catalog has been
examined in such great detail as the Abell catalog. The
discovery of line-of-sight anisotropies present in $R =0$ clusters is
a direct result of the catalog's detailed analysis (Sutherland 1988; Eftstahiou 1992;
Peacock \& West 1992).
These anisotropies in the $R=0$ clusters have led many researchers
to conclude that the correlation length of Abell clusters is artifically
enhanced and is not an accurate estimation of the scale of clustering
in the local Universe. A large correlation length with clustering on scales
$\sim 50h^{-1}$Mpc is not consistent with standard CDM models.
These results
provide continuing evidence for a large correlation length
for rich clusters, which must be represented in cosmological evolutionary
scenarios. 

\centerline{}
\noindent {\bf Acknowledgments}

\noindent{CM was funded in part by the National Aeronautics and Space Administration
and the Maine Science and Technology Foundation.
M.J.L was supported in part by NASA grant NAG5-6739.}

\begin{figure}
\epsfxsize=5.0in
\epsffile{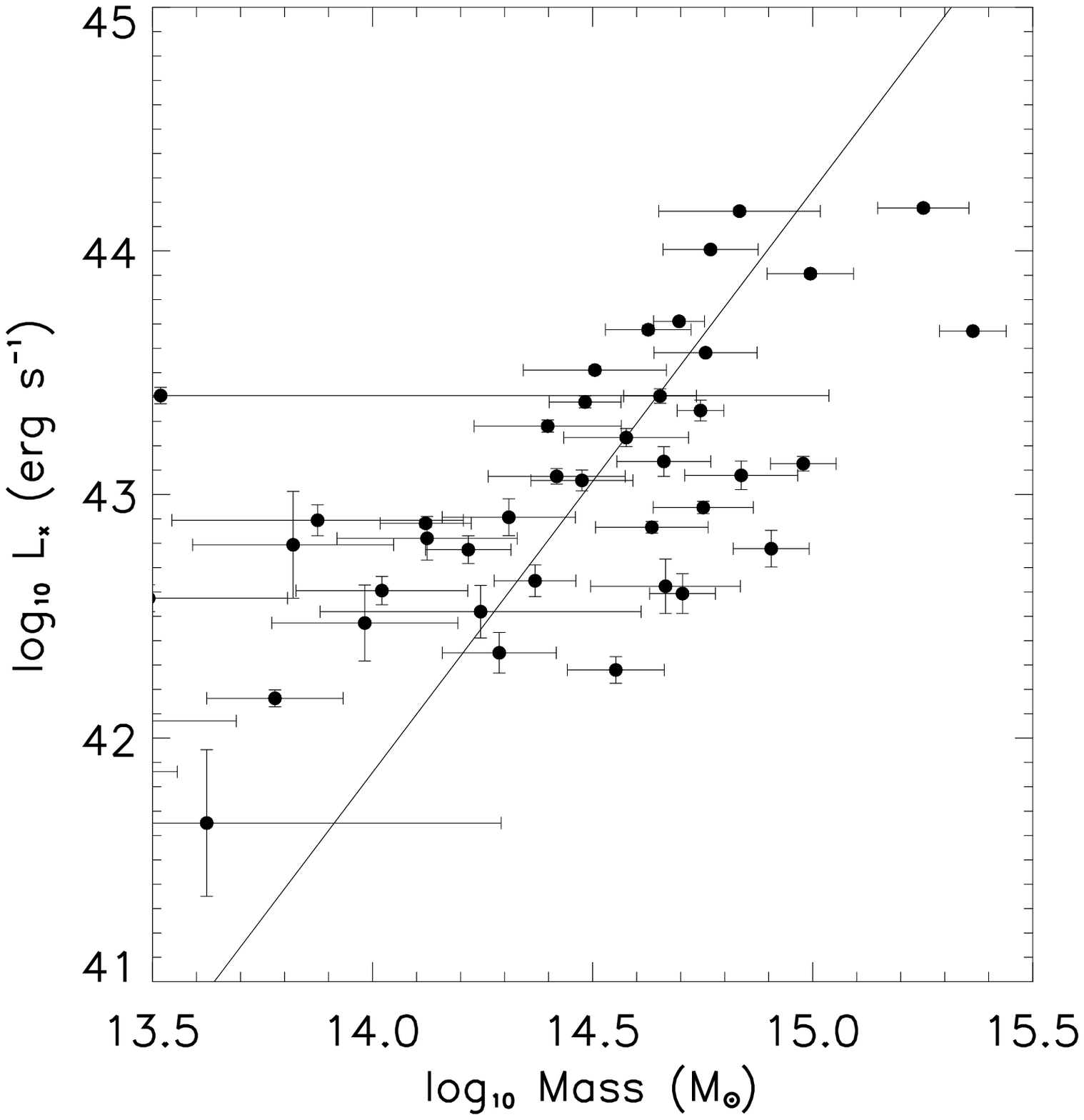}
\caption{ The $L_x - M$ relation for a volume-limited sample ($z \le 0.09$)
of X-ray bright Abell clusters.  The cluster virial masses were  published by
Girardi {\rm et al.} 1998. The line is an outlier-resistant, error-weighted
best fit with a slope of $2.38 \pm{1.3}$.}
\end{figure}

\newpage
\begin{figure}
\epsfxsize=5.0in
\epsffile{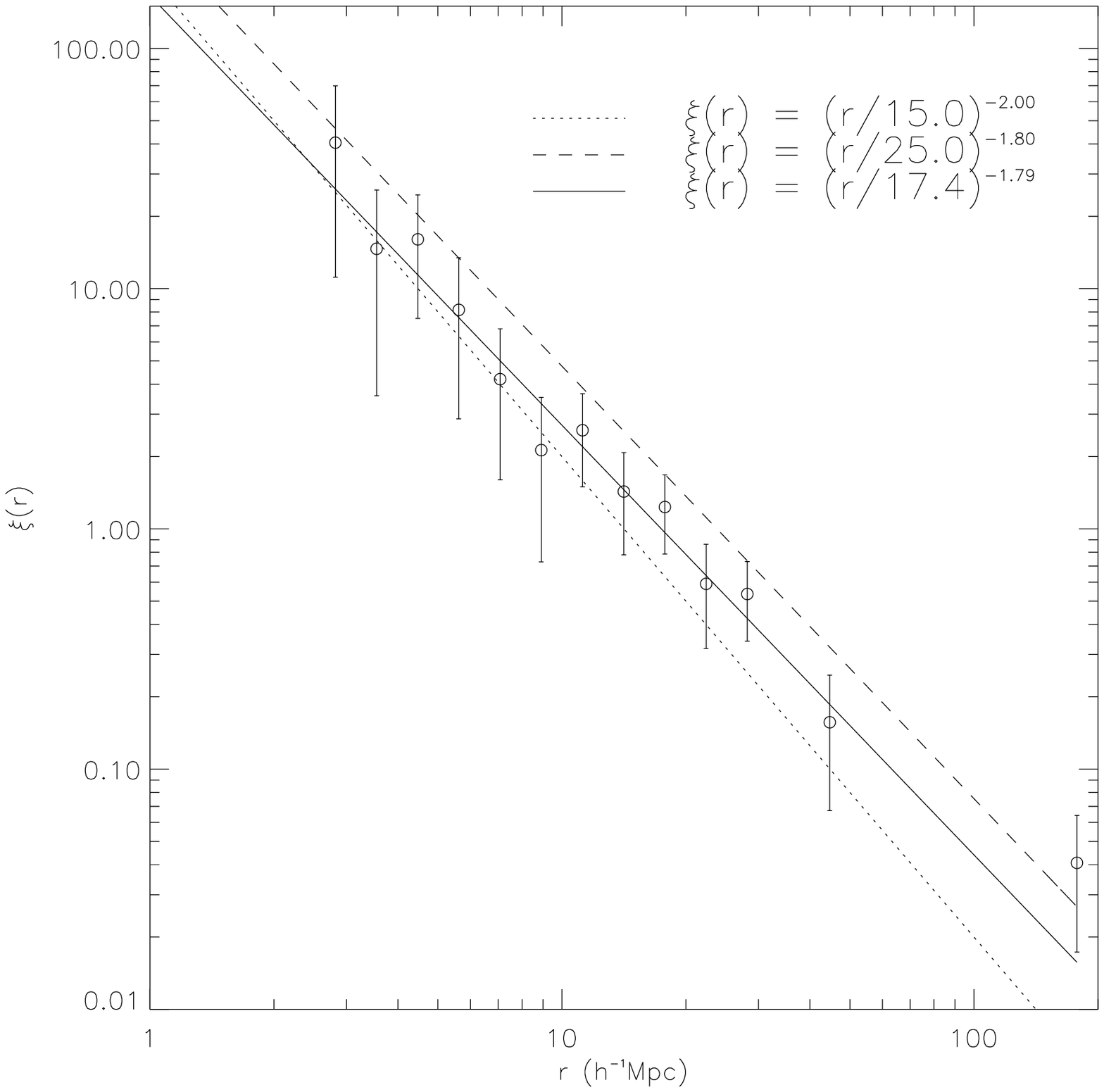}
\caption[]{$\xi(r)$ for the Voges {\rm et al} Sample 1 ($R \ge 0$ Abell clusters).
The dashed-line and dotted-line correspond to the two enveloping results
for $r_0$ and $\gamma$ from Bahcall \& Soneira (1983) (dashed) and
Efstathiou {\rm et al.} (1992) (dotted).}
\end{figure}

\begin{figure}
\epsfxsize=5.0in
\epsffile{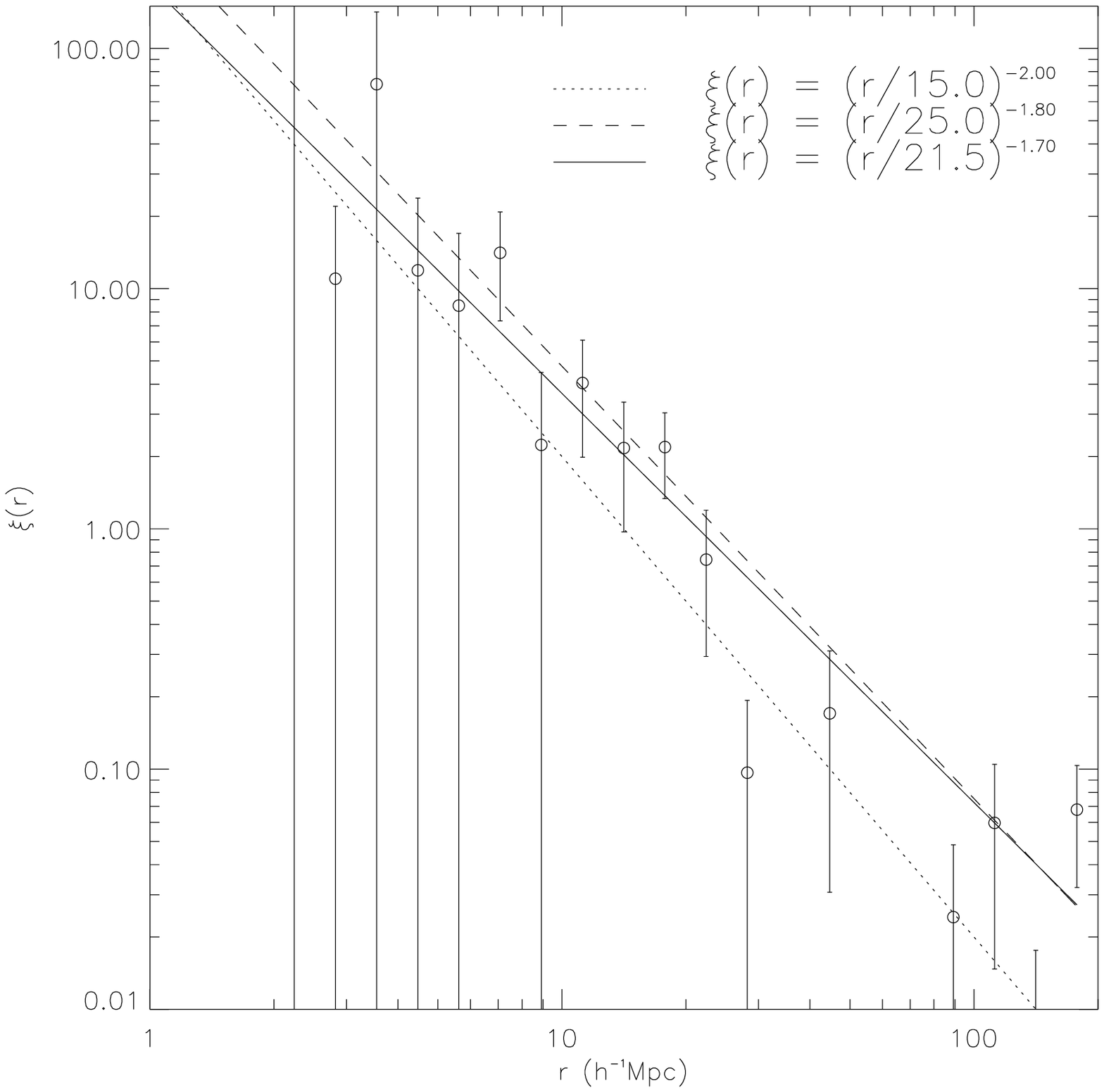}
\caption[]{$\xi(r)$ for the Voges {\rm et al.} Sample 3 ($R \ge 1$ Abell clusters).
The dashed-line and dotted-line correspond to the two enveloping results
for $r_0$ and $\gamma$ from Bahcall \& Soneira (1983) (dashed) and
Efstathiou {\rm et al.} (1992) (dotted).}
\end{figure}

\begin{figure}
\epsfxsize=5.0in
\epsffile{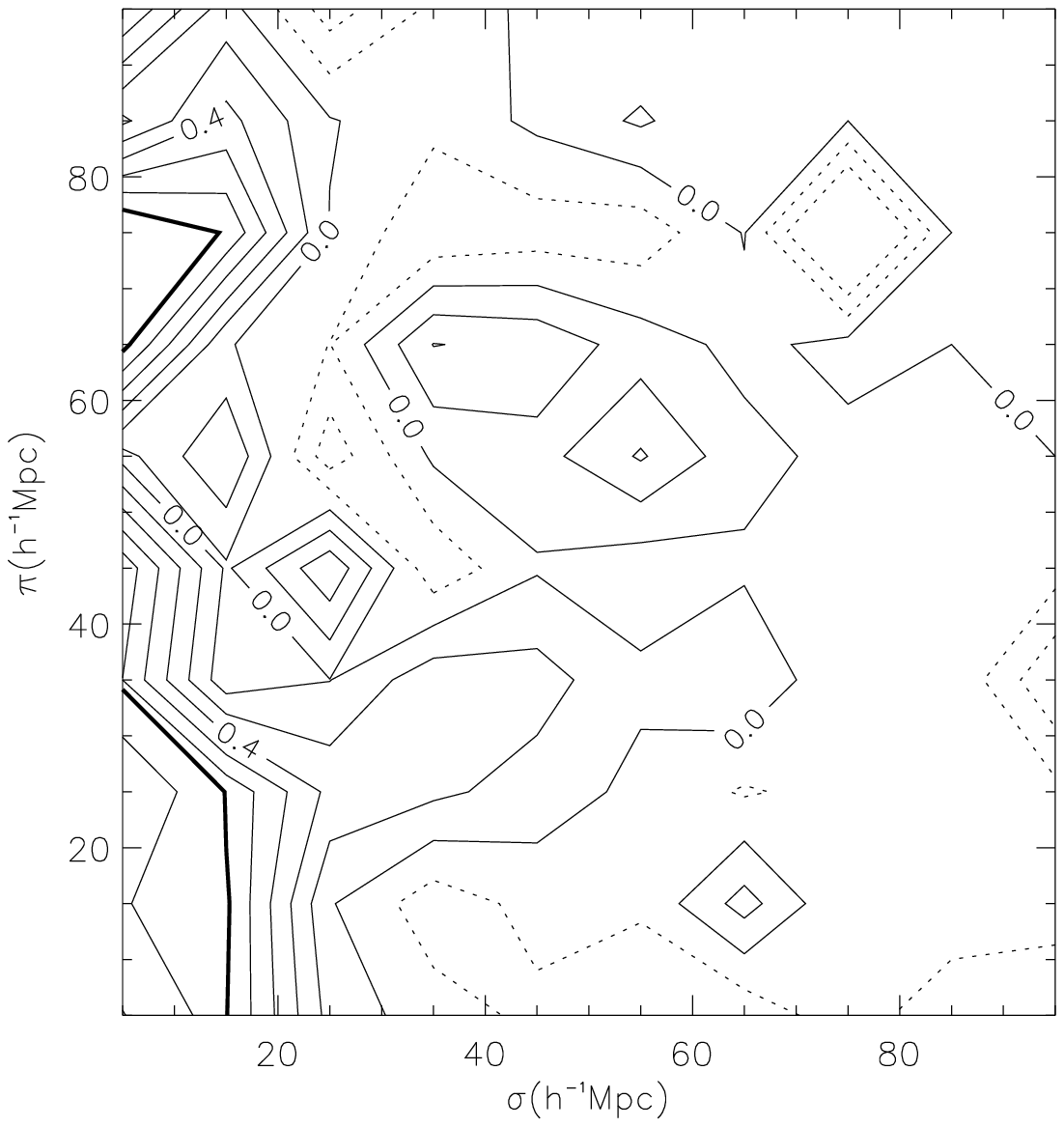}
\caption[] {
Contour plot of $\xi(\sigma, \pi)$ for the Voges {\rm et al.} Sample 1 ($R \ge 0$ Abell clusters).
The heavy contour corresponds to $\xi(\sigma, \pi) = 1$.}
\end{figure}

\begin{figure}
\epsfxsize=5.0in
\epsffile{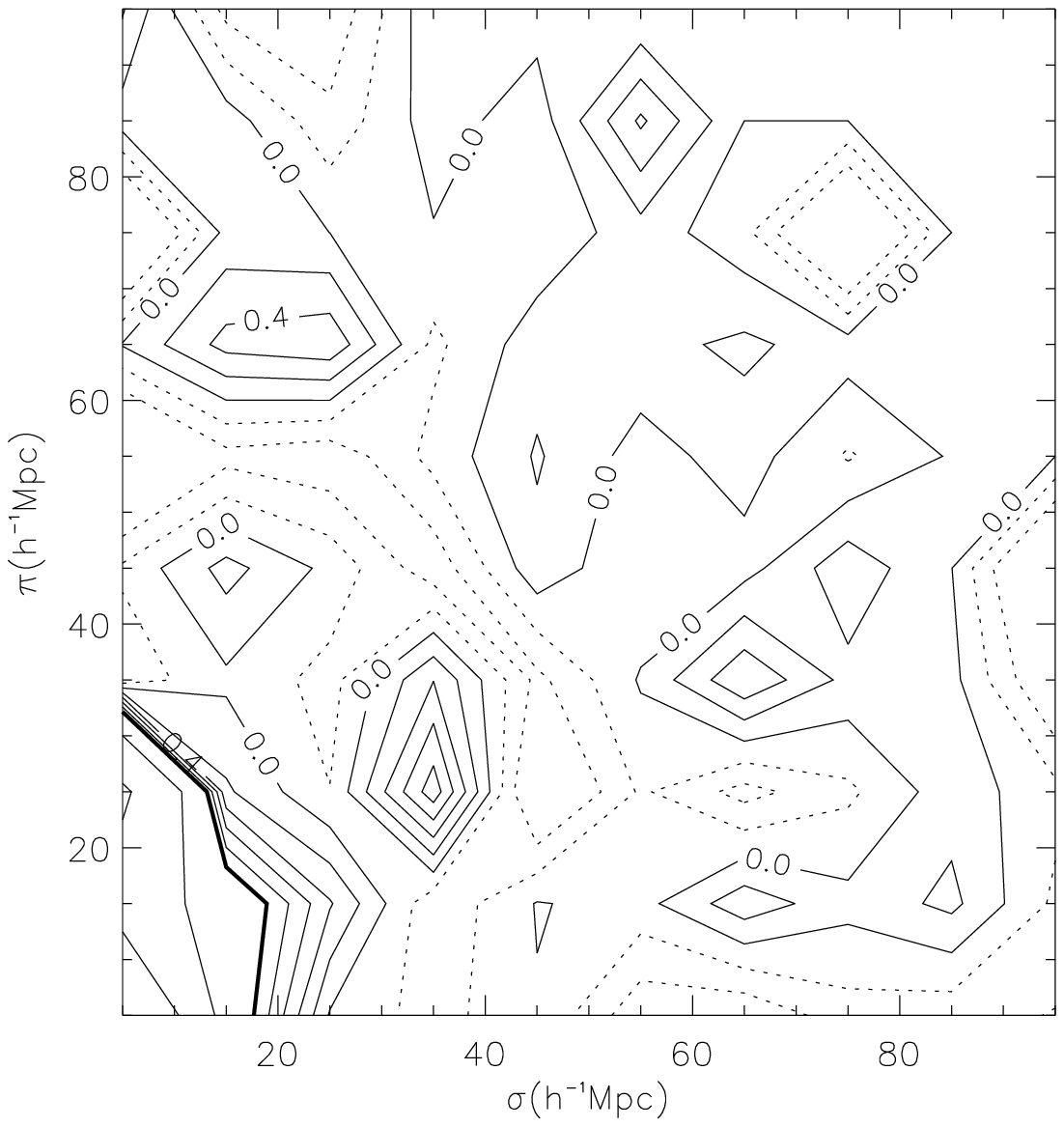}
\caption[] {
Contour plot of $\xi(\sigma, \pi)$ for the Voges {\rm et al.} Sample 3 ($R \ge 1$ Abell clusters).
The heavy contour corresponds to $\xi(\sigma, \pi) = 1$.}
\end{figure}

\begin{figure}
\epsfxsize=5.0in
\epsffile{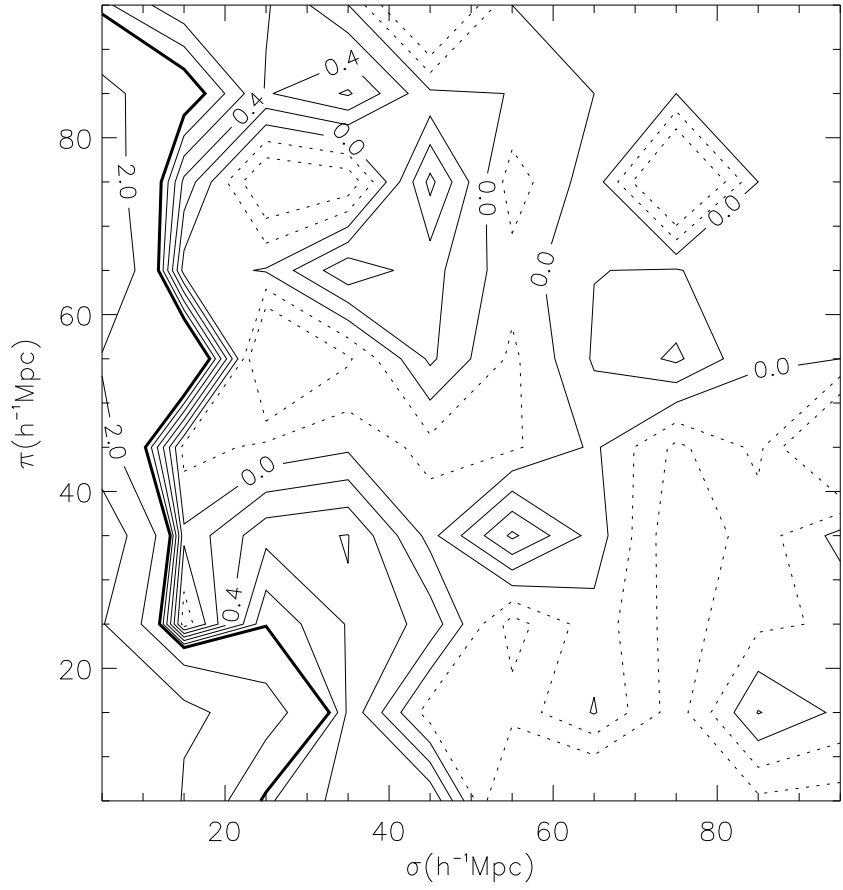}
\caption[] {
Contour plot of $\xi(\sigma, \pi)$ for the XBACs sample.
The heavy contour corresponds to $\xi(\sigma, \pi) = 1$.}
\end{figure}

\begin{figure}
\epsfxsize=5.0in
\epsffile{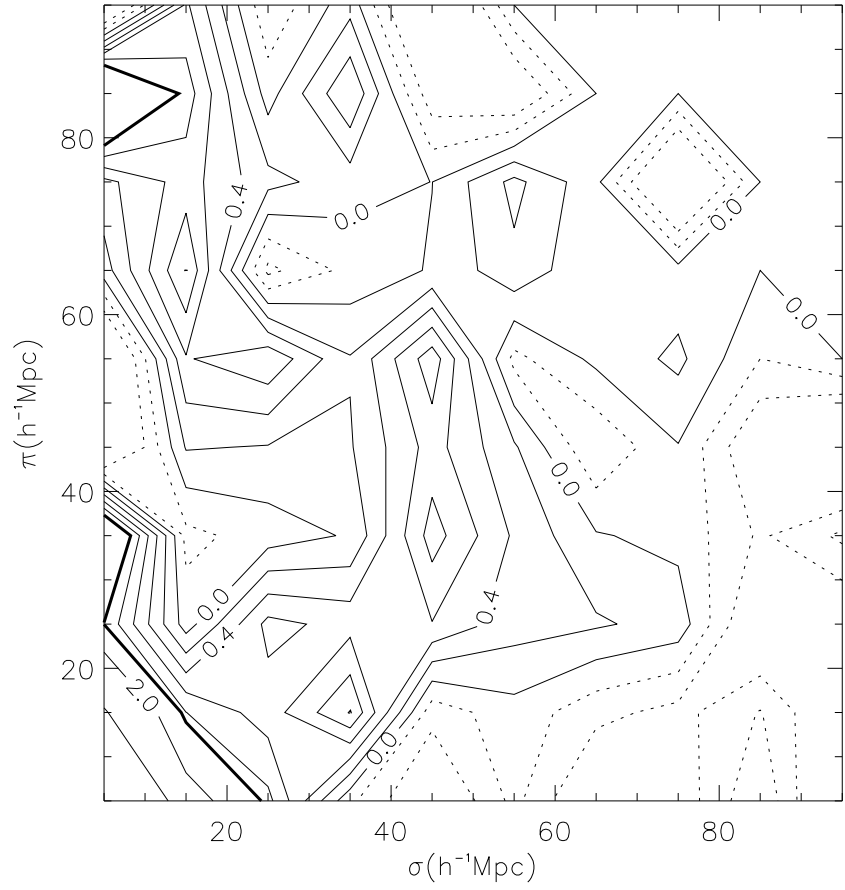}
\caption[] {
Contour plot of $\xi(\sigma, \pi)$ for the RASS1 clusters.
The heavy contour corresponds to $\xi(\sigma, \pi) = 1$.}
\end{figure}

\end{document}